\documentclass[twocolumn,aps,prl,floatfix,superscriptaddress]{revtex4-1}
\usepackage{amsmath,amssymb}
\usepackage{graphicx,float}
\usepackage{times,txfonts}
\usepackage{textcomp}
\usepackage{color}
\usepackage{multirow}
\usepackage{color}
\usepackage{soul}
\usepackage{hyperref}
\usepackage{natbib}
\usepackage{nicefrac}
\usepackage{multirow}
\usepackage{blindtext}
\usepackage[export]{adjustbox}

\renewcommand{\epsilon}{\varepsilon}

\usepackage{sistyle}

\usepackage{soul}
\definecolor{hugoColor}{RGB}{59,134,255}
\definecolor{YellowOrange}{RGB}{226,154,2}
\sethlcolor{YellowOrange}
\sethlcolor{lightblue}

\sethlcolor{lightblue}

\begin{document}
\title{Beam Steering with Ultracompact and Low-Power Silicon Resonator Phase Shifters}

\author{Hugo~Larocque}
\affiliation{Raytheon BBN Technologies, Cambridge, Massachusetts 02138, USA}
\affiliation{Research Laboratory of Electronics \& Department of Electrical Engineering and Computer Science, Massachusetts Institute of Technology, Cambridge, Massachusetts 02139, USA}
\author{Leonardo~Ranzani}
\affiliation{Raytheon BBN Technologies, Cambridge, Massachusetts 02138, USA}
\author{James~Leatham}
\affiliation{Raytheon Space and Airborne Systems, El Segundo, California 90245, USA}
\author{Jeffrey~Tate}
\affiliation{Raytheon Space and Airborne Systems, El Segundo, California 90245, USA}
\author{Alex~Niechayev}
\affiliation{Raytheon Space and Airborne Systems, El Segundo, California 90245, USA}
\author{Thomas~Yengst}
\affiliation{Raytheon Space and Airborne Systems, El Segundo, California 90245, USA}
\author{Tin~Komljenovic}
\affiliation{Nexus Photonics, Santa Barbara, California 93105, USA}
\author{Charley~Fodran}
\affiliation{Raytheon Space and Airborne Systems, El Segundo, California 90245, USA}
\author{Duane~Smith}
\affiliation{Raytheon Space and Airborne Systems, El Segundo, California 90245, USA}
\author{Mohammad~Soltani}
\affiliation{Raytheon BBN Technologies, Cambridge, Massachusetts 02138, USA}
\email{mo.soltani@raytheon.com}
%
\begin{abstract}
Photonic integrated circuit (PIC) phased arrays can be an enabling technology for a broad range of applications including free-space laser communications on compact moving platforms. However, scaling PIC phased arrays to a large number of array elements is limited by the large size and high power consumption of individual phase shifters used for beam steering. In this paper, we demonstrate silicon PIC phased array beam steering based on thermally tuned ultracompact microring resonator phase shifters with a radius of a few microns. These resonators integrated with micro-heaters are designed to be strongly coupled to an external waveguide, thereby providing a large and adjustable phase shift with a small residual amplitude modulation while consuming an average power of 0.4~mW. We also introduce characterization techniques for the calibration of resonator phase shifters in the phased array. With such compact phase shifters and our calibration techniques, we demonstrate beam steering with a 1x8 PIC phased array. The small size of these resonator phase shifters will enable low-power and ultra-large scale PIC phased arrays for long distance laser communication systems.    
\end{abstract}
\pacs{Valid PACS appear here}
\maketitle

\section{Introduction}

Next generation compact space and aerial moving platforms will highly rely on free-space laser communication systems deploying photonic integrated circuit (PIC) technology. A key enabling PIC component in this regard
is an integrated optical phased array~\cite{doylend2011two,sun2013large}, which has also found numerous applications ranging from chip scale light detection and ranging (LiDAR)~\cite{poulton2017coherent,poulton2019long} to beam shaping~\cite{notaros2017integrated}. State-of-the-art silicon photonics has particularly distinguished itself among other PIC platforms as a scalable means for reliably manufacturing phased-arrays and integrating them within modern technologies. Numerous silicon PIC design variations have been introduced to increase the performance of phased-arrays by implementing arrays with a larger number of elements ~\cite{sun2013large,sun2013large2,poulton2017large}, a wider beam steering range~\cite{yaacobi2014integrated,zadka2018chip,phare2018silicon,zhang2019sub}, and narrower beams in the far field~\cite{sun2014two,hutchison2016high,chung2017monolithically, fatemi2019nonuniform}. 

Despite many advances, one main limitation that prevents further scaling of PIC phased arrays is the size and power consumption of phase shifters. In particular, phase shifters which are typically waveguide-based and use the thermo-optic effect for phase shifting have a length scale of a millimeter, and a power consumption of several to tens of milliwatts~\cite{hutchison2016high,chung2017monolithically,phare2018silicon, fatemi2019nonuniform}. A waveguide-based phase shifter can be reduced to the micron scale, but at the expense of much higher electric power consumption, e.g. near ten milliwatts per phase shifter~\cite{sun2013large}. Alternative phase shifting approaches, including one based on light-recycling waveguides~\cite{miller2018512} and thermo-optic effects, have shown to dramatically reduce power consumption to a few milliwatts, though this level of power is still high and the phase shifter size is of the order of several hundreds of microns. To circumvent the issue of large power consumption, free-carrier based waveguide phase shifters have been demonstrated~\cite{poulton2019long} as an alternative to thermo-optic phase shifters with record microwatt scale electric power consumption. However, their large length still hinders scalability. Other waveguide phase shifter approaches include those based on the heterogeneous integration of indium phosphide on silicon~\cite{xie2019heterogeneous} which provides a relatively compact and low power alternative, but is still too large to address the issue of scalability. The latter is particularly relevant in two-dimensional phased array devices wherein each radiating element needs to have its own individual phase shifter, hence the need for further alternative optical phase shifters that simultaneously provide low power consumption and a smaller size to enable scalable phased arrays. 

Here, we propose and implement silicon PIC phased arrays that rely on ultra compact ring resonators as phase shifters. High quality factor ($Q$) silicon microring resonators with a radius within a few microns can provide compact phase shifters given that they can exhibit strong phase accumulation due to the large amount of round-trips experienced by light in the resonator near the resonant frequency. Resonator phase shifters have been demonstrated in silicon nitride integrated photonics~\cite{liang2019efficient}, where they occupy a larger footprint and consume larger powers. Electric tuning of our small footprint silicon resonators via thermo-optics, as demonstrated here, or free-carrier modulation~\cite{timurdogan2014ultralow} can induce large phase shifts with low power consumption. We design these resonator phase shifters to be strongly coupled to their external coupling waveguide in order to reduce any residual amplitude distortion near resonance. Our approach allows us to produce a device where each emitting unit cell of the array, which includes a radiating element and its corresponding phase shifter, occupies an area of $15 \times 20 \text{ \textmu m}^2$ and consumes an average power $0.4$~mW. We also introduce characterization procedures for these individual radiating phased elements based both on near-field and far-field imaging of the array. Using these resonator phase shifters, we demonstrate beam steering with a 1x8 phased array.

\section{Theory}

\begin{figure}[htbp]
\centering\includegraphics[width=\columnwidth]{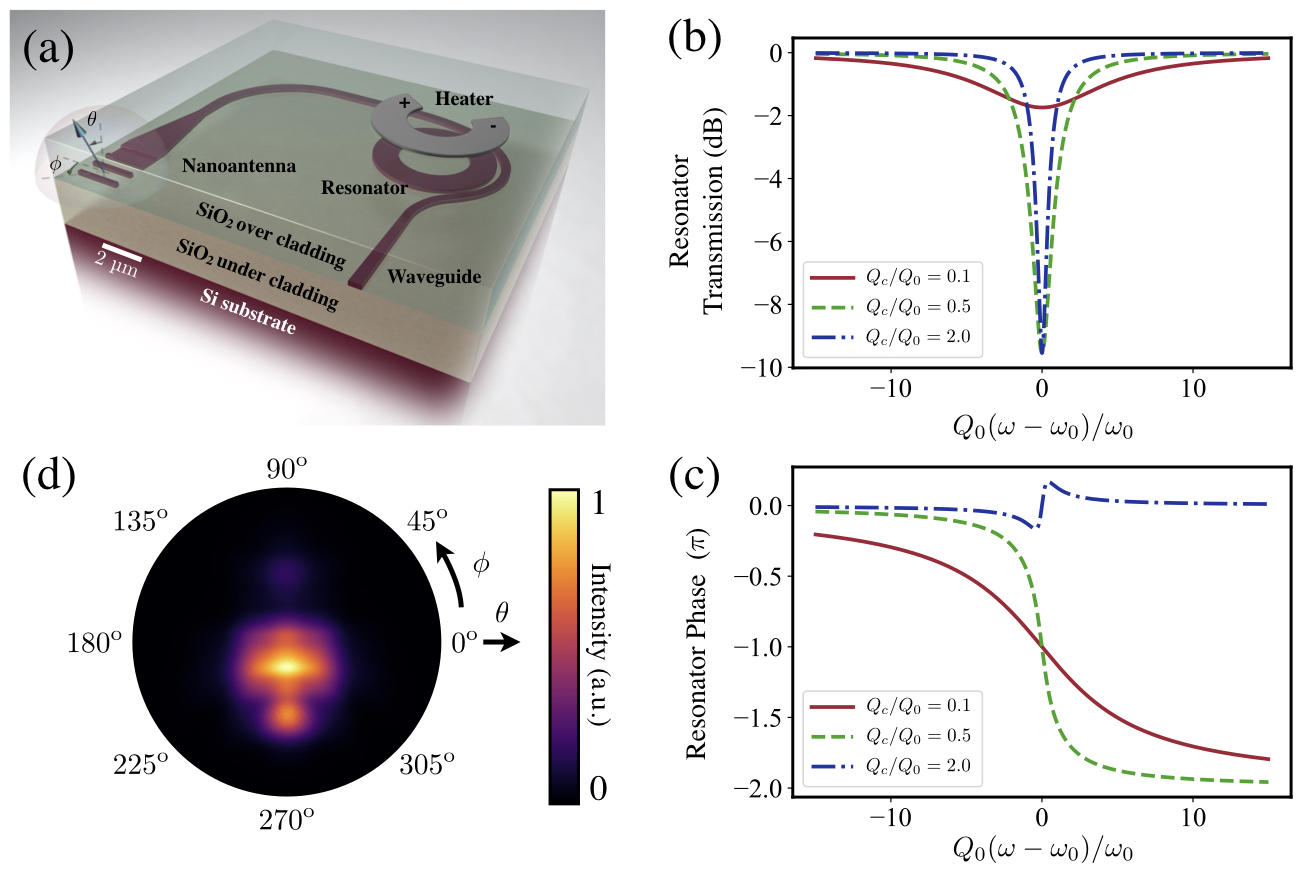}
\caption{\textbf{Design of the phased array unit cell.} (a) Three-dimensional representation of the PIC phased array unit cell where a silicon waveguide is coupled to a resonator tuned by a resistive heater. The resonator's output is then emitted into free-space with an optical nanoantenna. (b) Amplitude and (c) phase of the resonator transmission spectrum associated with three different waveguide-resonator coupling regimes attributed to over-coupled ($Q_c =0.5 Q_0$) and under-coupled ($Q_c =2 Q_0$), as well as strongly over-coupled ($Q_c =0.1 Q_0$) resonators. (d) Simulated far-field radiation pattern associated with the antenna design.}
\label{fig:theory}
\end{figure}

An illustration of a unit cell element of our PIC phased array is depicted in Fig.~\ref{fig:theory}(a) which includes a waveguide coupled to an optical resonator phase shifter placed under a resistive heater for phase tuning. The waveguide is then routed to an optical nanoantenna which emits the optical field into free-space.

A resonator's ability to impart large phase shifts with minimal amplitude distortion is highly dependent on its coupling to an external waveguide. This can be analyzed through the coupled-mode theory of waveguide-resonator coupling, according to which~\cite{soltani2010systematic}, the transmission of a resonator can be described by Eq.~(\ref{eq:resTransmission})
\begin{equation}
    \label{eq:resTransmission}
    T = \frac{j2(\omega-\omega_0)/\omega_0 + 1/Q_0 - 1/Q_c}{j2(\omega-\omega_0)/\omega_0 + 1/Q_0 + 1/Q_c},
\end{equation}
where $\omega$ and $\omega_0$ are the laser frequency and  the resonator's resonant frequency, respectively, $Q_0$ is the intrinsic quality factor of the resonator, and $Q_c$ is the coupling quality factor between the resonator and the waveguide. The shape of this transmission curve consists of a Lorentzian with a full width at half maximum (FWHM) given by the \textit{loaded} $Q$ of the resonator $Q_L$ such that $Q_L^{-1} = Q_0^{-1} + Q_c^{-1}$. To provide a substantial phase shift with small amplitude distortion, the PIC resonators must be strongly over-coupled, i.e. $Q_0 \gg Q_c$. This is verified in Fig.~\ref{fig:theory}(b,c) wherein control over an over-coupled resonator's $\omega_0$ can provide sharp variations in the phase of its transmitted field while maintaining minimal variations in its amplitude, thereby allowing such structures to be used as phase shifters. When $Q_0 \gg Q_c$, the phase of Eq.~(\ref{eq:resTransmission}) can be approximated as
\begin{equation}
    \label{eq:resPhase}
    \phi\approx-2\tan^{-1}\left(\frac{2Q_c(\omega-\omega_0)}{\omega_0}\right),
\end{equation}
which clearly shows how phase varies with the resonant frequency that can be shifted by altering the resonator refractive index. Here, we use the thermo-optic effect in silicon to produce these variations by means of the resistive heater placed over the resonator.

The nanoantenna emitter shown in Fig.~\ref{fig:theory}(a) features a standard antenna design~\cite{sun2013large,supp} that was optimized to achieve a large field of view as well as a large upward radiation efficiency. Fig.~\ref{fig:theory}(d) shows the simulated far-field radiation pattern of the antenna (see the supplementary material for more details concerning the nanoantenna structure~\cite{supp}).

\section{Methods}

\subsection{Device and Apparatus}

\begin{figure}[htbp]
\centering\includegraphics[width=\columnwidth]{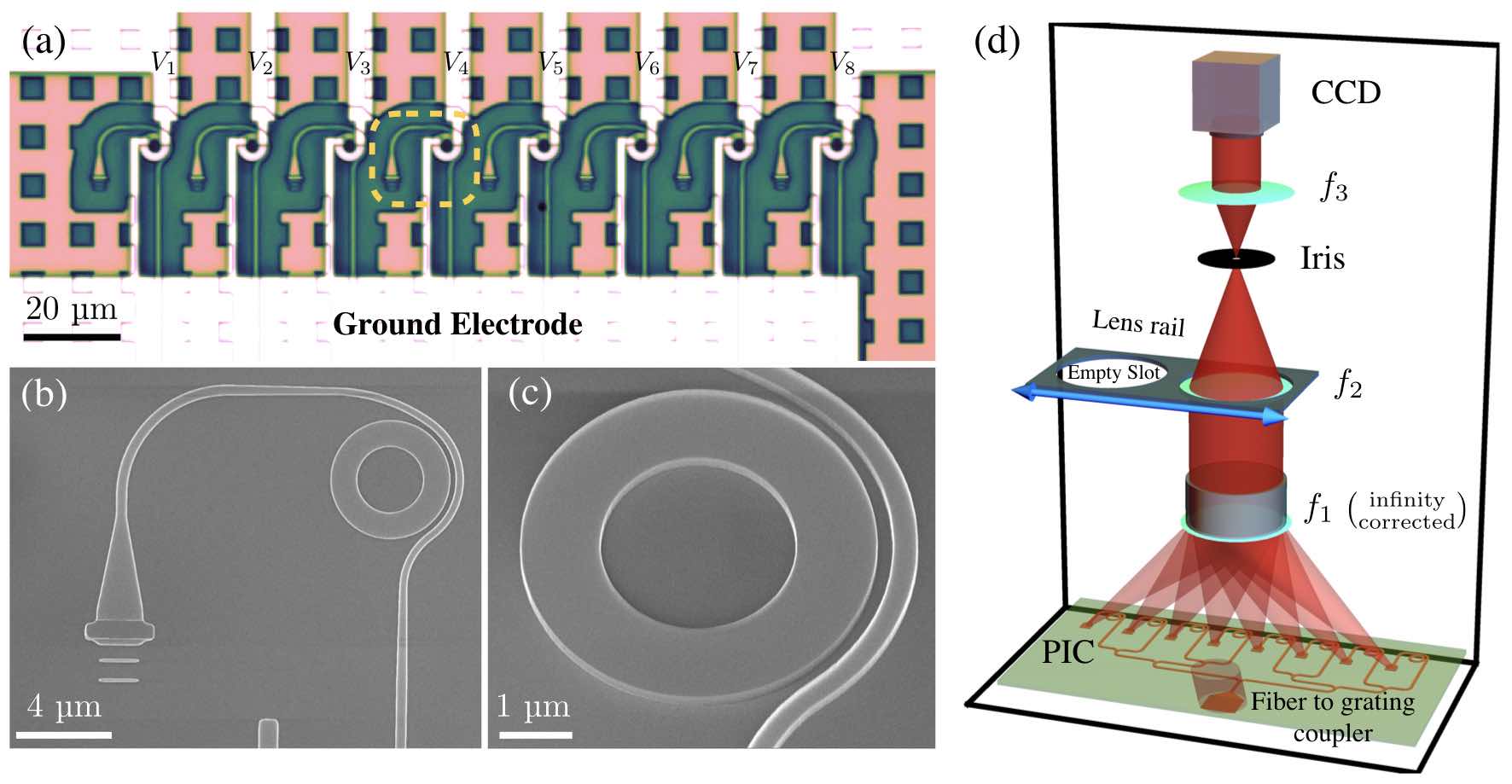}
\caption{\textbf{Chip layout and experimental apparatus.} (a) Optical micrograph of a 1x8 phased-array with resonator phase shifters displayed in false colors where the device ``unit cell'' is framed. (b) SEM image of the device's unit cell and (c) its over-coupled resonator. (d) Experimental apparatus with two switchable configurations for near-field and far-field imaging used to characterize the fabricated phased-arrays. Figure legend: PIC: photonic integrated circuit, $f_1$: objective lens, $f_2$: removable lens, $f_3$: imaging lens, CCD: charge-coupled device camera.}
\label{fig:chipApparatus}
\end{figure}

We implement a 1x8 PIC phased array with the array unit cell outlined in Fig.~\ref{fig:theory}(a) on a silicon photonic platform. Grating couplers are used to couple infrared light into the chip via an optical fiber. Fig.~\ref{fig:chipApparatus}(a) shows a micrograph of the device's active regions. We use compact and low loss splitters~\cite{zhang2013compact} to branch out laser power to the nanoantennas. The device is fabricated on a silicon-on-insulator wafer with a Si thickness of 220 nm on a 2~\textmu m SiO2 BOX layer. Using two steps of deep UV (248 nm) lithography and plasma etching, we define the photonic device layer components including waveguides, resonators, nanoantenna elements, and grating couplers. Fig.~\ref{fig:chipApparatus}(b) shows an SEM image of the unit cell with the resonator phase shifter and nanoantenna before cladding it with SiO2. As shown in Fig.~\ref{fig:chipApparatus}(c), the ring resonators with an external radius of 2.75~\textmu m and a width of 1.15~\textmu m are coupled to a waveguide with a pulley scheme~\cite{hosseini2010systematic} to provide strong over coupling. In the coupling region, the waveguide width is tapered down from 470~nm to 350~nm for better phase matching between the resonator mode and the waveguide mode. The gap between the waveguide and resonator is 240~nm. Since the minimum achievable waveguide-resonator gap for this fabrication was 240~nm, we used a relatively large pulley coupling angle to compensate for the weak waveguide-resonator mode overlap. The final device is cladded with 1.5~\textmu m SiO2 followed by the deposition of nichrome microheaters and gold contacts.

For the experiment, a micro-probe laid on gold pads on the chip is used to apply voltage to the resistive heaters of the resonators. We then characterize the performance of the device with a flexible imaging apparatus, shown in Fig.~\ref{fig:chipApparatus}(d), which can interchangeably measure both the near- and the far-field  of the phased array. Namely, the light emitted by the nanoantenna elements of the array is collected with a 20x infinity-corrected microscope objective lens (Edmund Optics, NA~=~0.6). The beam is thereafter focused onto an infrared CCD camera with a 12.5~cm lens to image the device's near-field. An additional 10~cm lens can be inserted into the imaging apparatus with a lens rail to convert the imaging lens into a 4f system imaging the device's far-field.

The ring resonator phase shifters designed for the phased array supports two radial modes (see supplementary material~\cite{supp}). The first order mode shows weak coupling to the external waveguide while the 2nd order mode shows strong over coupling. We characterize the $Q$ factor and the resonance wavelengths of these modes through test structure resonator devices with input/output fiber coupling and different waveguide-resonator coupling strengths fabricated on the same chip. We measure intrinsic $Q$'s of ${\sim} 19500$ and ${\sim} 6000$ for the 1st order and the 2nd order resonance modes. We use the 2nd order radial mode of the resonator in the phased array for beam steering, as it shows strong overcoupling to the waveguide. 

\subsection{Phased Array Calibration}

Though the resonator phase shifters were designed to be identical, fabrication imperfections often lead to slight discrepancies in their transmissive properties. The most important variations are in the resonant frequency, since the accumulated phase shift is most sensitive near this frequency. To align the resonances of the phase shifters, we apply a fixed bias voltage to their corresponding heaters. Once these voltages are applied, we employ a genetic algorithm to find the voltage values that optimize the amount of light steered to a given position on the camera~\cite{hutchison2016high}. These new voltages are then saved into a look-up table to be accessed in later experiments involving beam-steering.

We employ two methods based on near-field and far-field imaging to characterize the spectrum of the resonator phase shifters. The near-field method is initially used to learn the location of the resonance wavelengths, the resonance transmission extinction, and to perform a coarse alignment of the resonances. Though our near-field method is primarily applicable to resonances with observable extinction in their spectrum, the alignment of these resonances automatically align the other resonances including the strongly overcoupled ones. The far-field method is then used for the fine alignment of the strongly overcoupled resonance modes which are the modes of interest, and by nature have weak resonance extinction. 

\begin{figure}[t]
\centering\includegraphics[width=\columnwidth]{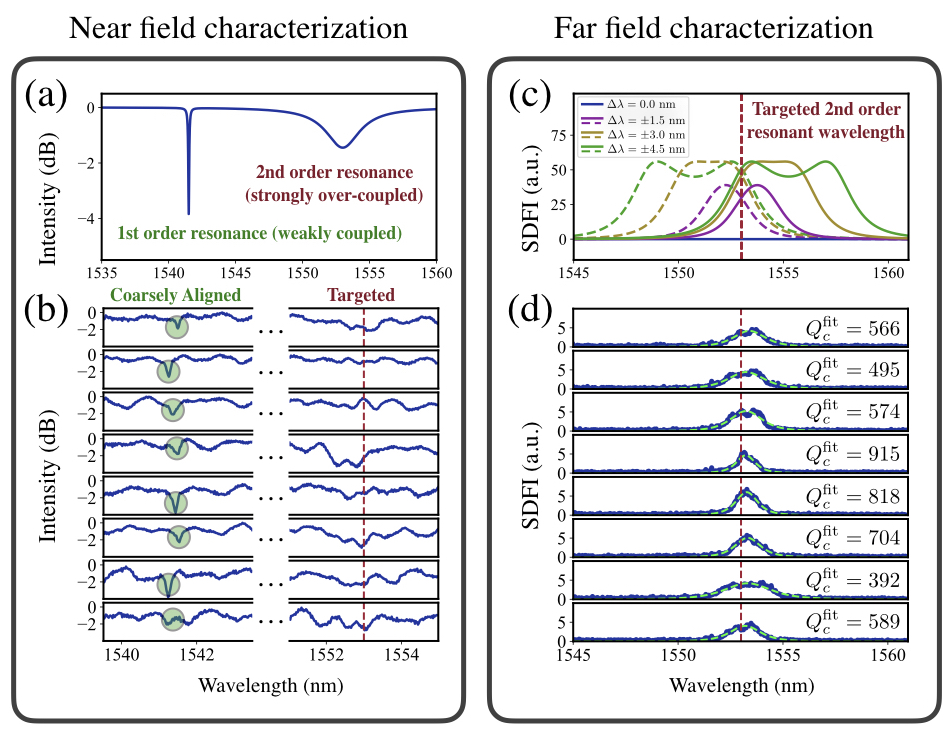}
\caption{\textbf{Resonator phase shifter characterization for the phased array.} (a) A representative resonator spectra based on Eq.~(\ref{eq:resTransmission}) expected from our resonator phase shifter, when using our near-field method. The first order mode is weakly coupled, and the 2nd order mode is strongly over-coupled, which is the case of interest . (b) Near-field measured spectrum of each resonator phase shifter in the phased array. The resonances are already aligned through an applied bias voltage to the heaters while monitoring the 1st order resonance which has an observable extinction. (c) Representative SDFI  spectrum of a resonator phase shifter based on Eq.~(\ref{eq:farfieldSpectrum}) in the phased array for different resonance shifts. (d) Measured SDFI spectrum of each resonator phase shifter in the array from Fig.~\ref{fig:chipApparatus}(a). The insets show the extracted $Q_c$ of the resonators via theoretical fitting of Eq.~(\ref{eq:farfieldSpectrum}).}
\label{fig:spectrum}
\end{figure}

The near-field method involves performing a wavelength sweep in which near-field images of the phased array output are collected by an IR camera. By reading the camera pixel intensities corresponding to individual nanoantennas at a given wavelength, we find the spectrum of the resonator phase shifter corresponding to each nanoantenna (see supplementary material for more details~\cite{supp}). Figure~\ref{fig:spectrum}(a) shows a representative spectrum of what we expect to achieve from such a measurement, and Fig.~\ref{fig:spectrum}(b) shows the measured spectrum of each resonator phase shifter in the array. As seen from this figure, though the near-field method is able to find the resonance attributed to the first order resonance mode, it cannot reliably locate and characterize the features of the 2nd order mode which is strongly over-coupled and exhibits minute transmitted intensity variations across its resonance. Nevertheless, by using this method to align the 1st order modes that have a noticeable extinction, the 2nd order mode becomes coarsely aligned. 

To overcome the limitations of the near-field method for the characterization and calibration of strongly overcoupled resonances, we introduce a far-field method that consists of monitoring the far-field image of the phased array whilst tuning the resonances of the phase shifters. The far-field method is primarily useful for overcoupled resonance modes that provide a large phase shift, and therefore a noticeable beam displacement in the far field. For a given wavelength, we monitor the dissimilarity between far-field images obtained when non-zero and zero bias voltages are applied to a single resonator phase shifter in the array. To quantify this dissimilarity, we calculate the squared differential far-field intensity (SDFI) integrated over the image plane for these two voltages, i.e. $\int(I_2(V,\lambda)-I_1(0,\lambda))^2 \,dx\,dy$, wherein $I_2$ and $I_1$ are the far-field image intensity patterns for applied bias voltages of $V$ and $0$, respectively. If the wavelength is close to resonance, larger phase shifts are expected to occur for a given applied voltage, thereby leading to far-field fringe displacements, hence a larger SDFI which allows to identify the resonance wavelength.

When the array is periodic, one can obtain the following analytical expression for the SDFI spectrum corresponding to each resonator phase shifter in the array (see supplementary material~\cite{supp} for its derivation):
\begin{equation}
    \label{eq:farfieldSpectrum}
    \begin{split}
 \text{SDFI}_n(\lambda) &= 16 \sin^2\left(\frac{\Delta\phi(\lambda)}{2}\right)\left[(N-1)\sin^2\left(\frac{\Delta\phi(\lambda)}{2}\right) \right. \\
 &\left. +\left\lvert\frac{2n-N-1}{2}\right\rvert\cos{(\Delta\phi(\lambda))}\right]
 \end{split}
\end{equation}
where $N$ is the total number of radiating nanoantenna elements in the array, $n$ is the index position of the considered antenna and its corresponding phase shifter, and $\Delta \phi (\lambda)$ is the phase shift obtained from applying a non-zero bias voltage $V$ to the $n^\text{th}$ resonator phase shifter, as expected from Eq.~(\ref{eq:resPhase}). The collected SDFI spectrum also allows for the extraction of the resonator properties by fitting it to Eq.~(\ref{eq:farfieldSpectrum}).  

Figure~\ref{fig:spectrum}(c) shows representative SDFI spectra of a resonator phase shifter in the phased array obtained from Eq.~(\ref{eq:farfieldSpectrum}) for various resonance shifts ($\Delta\lambda$) affecting $\Delta\phi$ as according to Eqs.~(\ref{eq:resTransmission},\ref{eq:resPhase}). For these plots, we consider a resonator mode with $Q_0$ and $Q_c$ values similar to what is expected in the 2nd order resonance mode of our fabricated device, i.e., values attributed to strong overcoupling with a resonance wavelength near 1553~nm. As seen from Fig.~\ref{fig:spectrum}(c), depending on the scale of the resonance shift $\Delta\lambda$, different SDFI spectral profiles are observed.    Figure~\ref{fig:spectrum}(d) shows the measured SDFI spectrum of each resonator phase shifter in the array, when two bias voltages of 0.2V and 0V are applied to each phase shifter, individually. The results match well with the predicted spectral profiles shown in Fig.~\ref{fig:spectrum}(c). The acquired SDFI spectra can then be used to extract the coupling quality factor of the resonator shifters given that it dominates the line-shape of strongly over-coupled resonators, i.e., for $Q_0 \gg Q_c$, $Q_L^{-1} \approx Q_c^{-1}$. Fits of the measured SDFI spectra to Eq.~(\ref{eq:farfieldSpectrum}) and their corresponding $Q_c$ values are displayed in Fig.~\ref{fig:spectrum}(d).

\section{Results}

\begin{figure}[htbp]
\centering\includegraphics[width=\columnwidth]{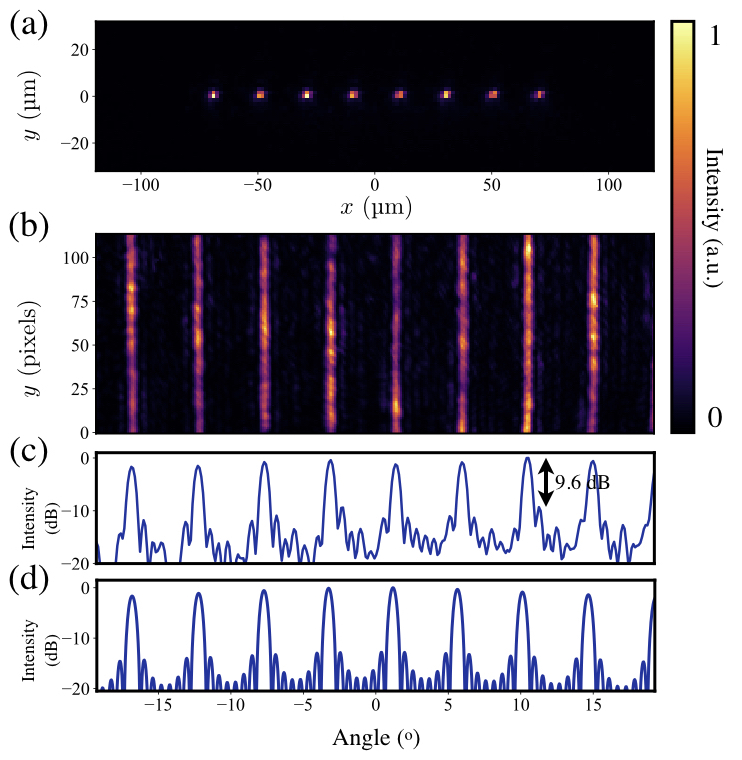}
\caption{\textbf{Near-field and optimized far-field patterns of a 1x8 phased array with resonator phase shifters shown in Fig.~\ref{fig:chipApparatus}(a).} (a) Near-field pattern of the phased array. (b) Far-field pattern of the phased-array optimized to have strong contrast between central peaks and their side-lobes. (c) Far-field pattern averaged along the $y$ direction. (d) Corresponding far-field pattern expected from theory.}
\label{fig:devicePattern}
\end{figure}

We proceed by applying the calibration method described in the previous section on the PIC phased array device shown in Fig.~\ref{fig:chipApparatus}. We first find the location of the resonance wavelengths using the test structures described in Fig.~\ref{fig:Spectra}~\cite{supp} as well as through the near-field characterization of the resonances in the phased array. With this information, we send laser light to the phased array with a wavelength near the expected resonance of the strongly overcoupled mode. The resulting far-field pattern of the array is then optimized to display high contrast features over a period determined by that of the nanoantennas' radiated field. These features are obtained for the optical wavelength to which the device resonance were calibrated, which in this case, consists of 1553~nm. The device's measured near-field is shown in Fig.~\ref{fig:devicePattern}(a) and its corresponding far-field fringe pattern can be found in Fig.~\ref{fig:devicePattern}(b) which shows a fringe period of $4.5^\text{o}$ as determined by the array pitch period of 20~\textmu m. Though this work focuses on periodically arranged unit cells producing periodically spaced fringes in the far-field, a random sparse array with a larger quantity of widely spaced radiating elements with respect to the unit cell size can produce a single lobe in the far-field~\cite{hutchison2016high,komljenovic2017sparse, fatemi2018scalable}. As displayed in Fig.~\ref{fig:devicePattern}(c), averaging the far-field pattern along the $y$ direction reveals that its main features have an extinction ratio of around 9.6 dB. For reference, we display the corresponding far-field pattern expected from theory in Fig.~\ref{fig:devicePattern}(d), which was obtained from the nano-antenna far-field pattern and the periodicity of the device unit cells.

\begin{figure}[htbp]
\centering\includegraphics[width=\columnwidth]{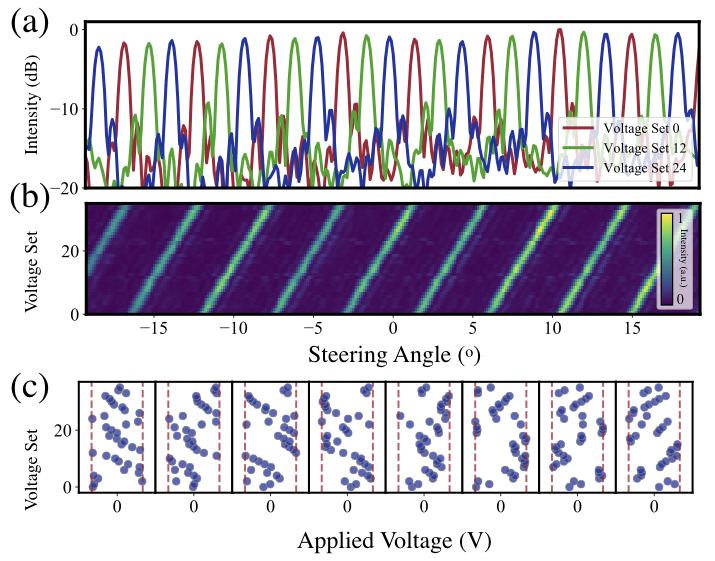}
\caption{\textbf{Beam steering with a 1x8 PIC phased array with resonator phase shifters.} (a) Beam-steering of the pattern shown in Fig.~\ref{fig:devicePattern} to three different positions. (b) Continuous beam-steering of the pattern shown in (a) achieved through the use of 36 different voltage configurations applied to the device's heaters. (c) Employed voltage values shifted with respect to the calibrating voltage used to shift the device resonances. $\pm$ 0.2 V regions are shown as red dotted lines.}
\label{fig:steering}
\end{figure}

As illustrated in Fig.~\ref{fig:steering}(a), the calibration procedure can be conducted to laterally shift this optimized far-field fringe pattern. With this method, we are able to steer the far-field pattern over one fringe period, i.e. $4.5^\text{o}$. An intensity plot of the steered beam cross-section is shown in Fig.~\ref{fig:steering}(b). Ignoring the voltage required to align each of the resonators' resonances, 0.2 V had to be applied to each heater on average to properly steer the beam while maintaining a decent amount of contrast between the pattern's main peaks and its side-lobes. These offsetted voltage values are displayed in Fig.~\ref{fig:steering}(c). Given that each of the device's heaters have a resistance of approximately 100~$\Omega$, then roughly 0.4 mW per channel is required to operate the device.

\section{Discussion}
Though in this work we used microheaters for thermo-optic tuning of resonator phase shifters, a more efficient alternative approach could involve charge injection phase shifters integrated with microresonators that can consume a much lower power~\cite{timurdogan2014ultralow}. The resonators fabricated in our devices were prone to fabrication errors and did not show identical resonances, and therefore required further tuning and static microheater power consumption to align the resonance wavelengths. For a range of devices tested in this work the static power consumption applied to phase shifters had a statistical distribution within the range of 0.1-3.6 mW. A better fabrication of these resonators in a trusted commercial foundry will provide more identical resonances and hence will dramatically reduce the static power consumption to align the resonators. For these devices we used the second order resonance modes of the resonator for the phase shifter, because it could provide a larger waveguide-resonator coupling strength for a waveguide-resonator gap of 240 nm limited by the resolution of our lithographic process. With higher resolution lithography and smaller waveguide-resonator gaps, one can use the first order resonance mode which has a higher $Q$, and design it in the strongly overcoupled regime with a smaller resonance linewidth, thereby allowing it to consume a lower amount of electric power for phase shifting. 

While in this work we demonstrated a one dimensional phased array, our concept can be extended to two dimensional arrays as well. A two dimensional (2D) phased array with an optimally randomized sparse placement of unit cells and appropriate phases can localize the beam in the far field with minimal side-lobes and enable a much larger steering range even if the array unit cell spacing is quite large with respect to the operation wavelength. Random sparse arrays of scatterers for field localization which date back to radar~\cite{lo1964mathematical,mailloux2017phased} and condensed matter physics~\cite{anderson1958absence}, have also been recently demonstrated in PIC phased array configurations~\cite{komljenovic2017sparse, fatemi2018scalable} and other optically random media~\cite{segev2013anderson}. Another challenge is the electric wiring of the resonator phase shifters within the 2D array which can be addressed by three dimensional integration of photonics and electronics.

\section{Conclusion}
In this work, we have demonstrated PIC phased array beam steering using ultra compact and low power silicon resonator phase shifters. We demonstrated a 1x8 phased array with an array unit cell size as small as 15 x 20 \textmu m$^2$ that includes a phase shifter with a power consumption as small as 0.4~mW for $\pi$ phase shifts. The compactness of the resonator phase shifter and the nanoantenna allows for the design of a compact and low power unit cell for phased arrays and provides a viable path for efficient and scalable 2D beam steering. We introduced new methods for the calibration of the resonator phase shifters in the array based on near-field and far-field information. More specifically, we developed a far-field analytical model for the spectral characterization of the strongly overcoupled resonator phase shifters in the array and used the model for the resonator characterization and the array calibration. Though we used thermo-optic tuning for the resonator phase shifters, employing charge injection phase shifters within the resonators and 3D integration of the PIC devices and electronics can enable even lower power an larger scale 2D beam steering platforms.

\section*{Acknowledgments}
The authors would like to thanks Dr. Stephen Palese from Raytheon Space \& Airborne Systems for helpful discussion and critical suggestions on this work. This document does not contain technology or technical data controlled under either the U.S. International Traffic in Arms Regulations or the U.S. Export Administration Regulations.

\bibliography{arrayBib}

\newpage

\section{Supplementary Material}

\subsection{Resonator Modes}

\begin{figure}[htbp]
\centering\includegraphics[width=\columnwidth]{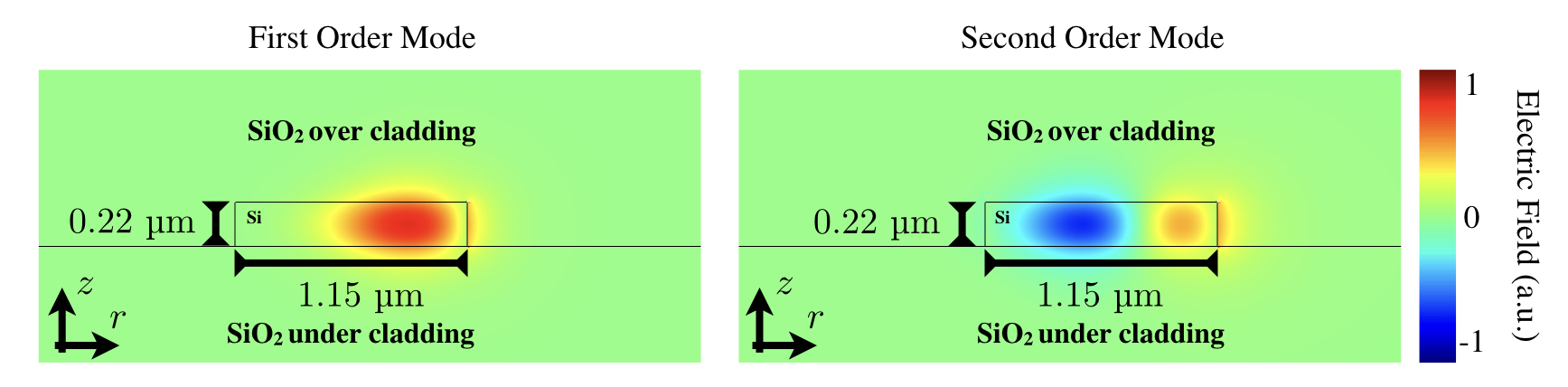}
\caption{\textbf{Transverse modes supported by the ring resonator.} }
\label{fig:modes}
\end{figure}

The microring resonator phase shifter used in this work has an external radius of 2.75 micron, a thickness of 220 nm, and a ring width of 1.15 micron. We employed the transverse electric (TE) resonance mode of the resonator for our design as it can be supported with a more compact resonator geometry. A finite element method is employed to find the resonance optical modes of the device. Figure~\ref{fig:modes} shows the 1st and the 2nd order radial TE mode profiles supported by the resonator. We use the 2nd order mode for the phase shifter operation due to its strongly over coupling to the external waveguide.

\subsection{Resonator Spectra and its Coupling Strength to a Waveguide}

\begin{figure}[t]
\centering\includegraphics[width=\columnwidth]{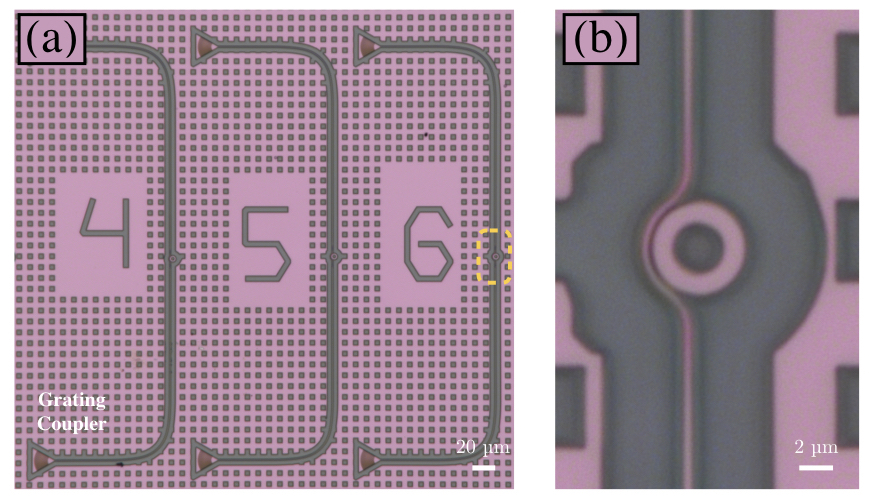}
\caption{\textbf{Test structure micrographs.} (a) Test structures with input and output grating couplers, and resonators with a radius of 2.75 \textmu m and pulley coupling angles of $60^\text{o}$ (4), $90^\text{o}$ (5), and $120^\text{o}$ (6). (b) Zoomed-in version of (a) focusing on the resonator employed as the phase shifter in the PIC phased array in the main text. The measured insertion loss of the grating couplers is 6-8 dB}
\label{fig:resonatorPic}
\end{figure}
 An efficient resonator phase shifter used in the phased array needs to be strongly overcoupled with weak extinction to minimize amplitude distortion. Hence, initially identifying and locating such a resonance from the background spectrum can be difficult. To simplify the resonance wavelength search as well as the systematic characterization of the resonator coupling strength to an external pulley waveguide, we fabricated a series of test structures adjacent to the PIC phased array device. Each test structure, shown in  Fig.~\ref{fig:resonatorPic}, consists of a resonator coupled to an external waveguide with a coupling gap of 240 nm and different pulley coupling angles, along with two grating couplers at both ends for input and output coupling to an optical fiber.

\begin{figure}[htbp]
\centering\includegraphics[width=\columnwidth]{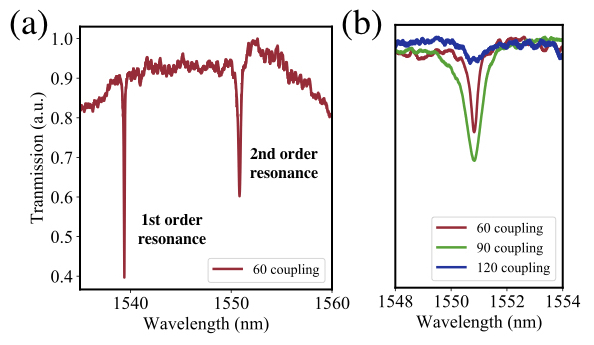}
\caption{\textbf{Transmission spectra of the ring resonators in the test structures shown in Fig.~\ref{fig:resonatorPic}}. (a) Spectrum of a resonator with $60^\text{o}$ pulley coupling featuring its first and second order modes at 1539.4~nm and 1550.8~nm, respectively. (b) Region of the resonator spectrum focusing on the second order mode for the cases of $60^\text{o}$ (under-coupled), $90^\text{o}$ (close to critically coupled), and $120^\text{o}$ (over-coupled) pulley coupling.}
\label{fig:Spectra}
\end{figure}

The spectra of these resonators for different pulley coupling angles are provided in Fig.~\ref{fig:Spectra}.
The first order resonances of the device are observed to be near 1535~nm and exhibit an excessive amount of extinction, thereby preventing their usefulness in our phased arrays. However, the 2nd order resonances observed near 1549~nm have a sufficiently low extinction ratio. For the 2nd order mode, we observe that resonators with a pulley coupling length of $60^\text{o}$ appear to be under-coupled, $90^\text{o}$ are closer to being critically coupled, and that devices with a $120^\text{o}$ coupling are over-coupled. For this reason, resonators with $120^\text{o}$ pulley coupling length, also shown in Fig.~\ref{fig:resonatorPic}(b), were chosen as the phase-shifting PIC component of our phased arrays. From the measured spectra in Fig. S3, the intrinsic $Q$'s of the 1st order and the 2nd order resonance modes are found to  be ${\sim}19500$ and ${\sim}6000$, respectively. 

Characterizing the resonator test devices in Fig.~\ref{fig:resonatorPic} and locating the resonance wavelength and its spectrum as shown in Fig.~\ref{fig:Spectra} help us with narrowing down the wavelength sweeping range for the actual resonator farfield characterization in the phased array. For example by looking at Fig.~\ref{fig:Spectra}, we learn that the strongly overcoupled 2nd order resonance mode is near 1550 nm, thereby we use this wavelength zone in the farfield resonator characterization.  

\subsection{Nanoantenna Design}

\begin{figure}[t]
\centering\includegraphics[width=\columnwidth]{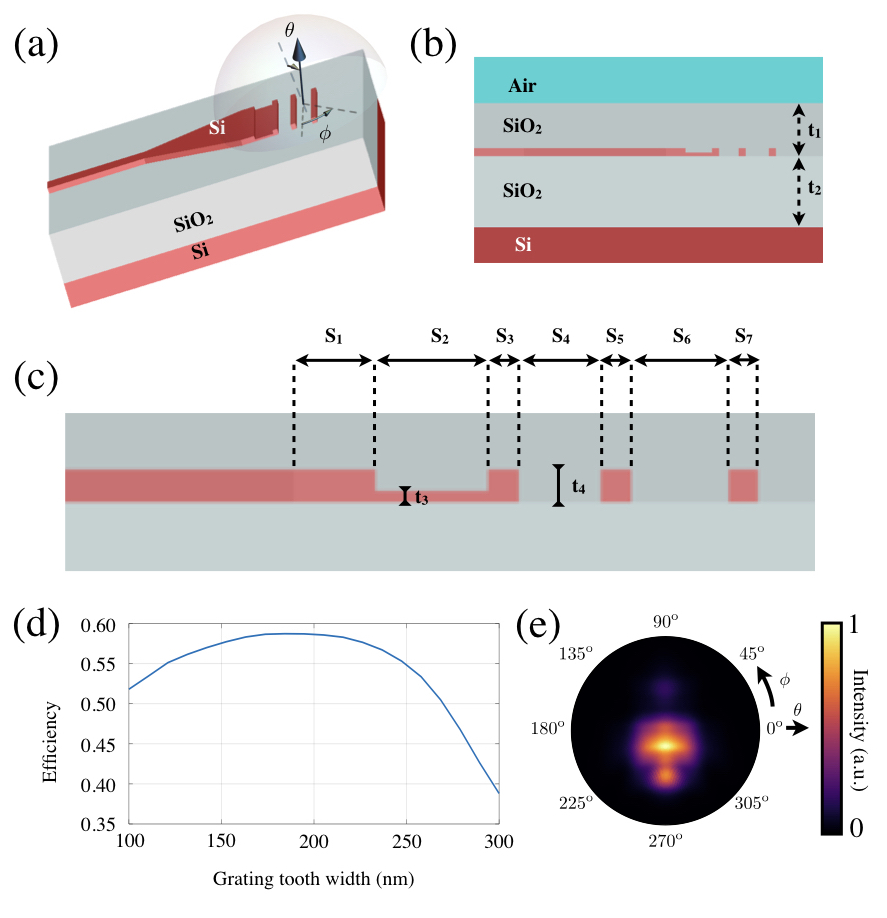}
\caption{\textbf{Antenna design employed in the phased array.} (a) Three-dimensional layout of the antenna. A 470 nm wide silicon on oxide waveguide tapers out over a distance of 4~\textmu m into the radiating component of the structure. (b) Two-dimensional cross-section of (a) highlighting the device structure where $t_1=$1.5 \textmu m  and $t_2=$2 \textmu m. (c) Zoomed-in version of (b) featuring the finer features of the antenna where $t_3=$140 nm, $t_4=$220 nm, $S_1=$550 nm, $S_2=$750 nm, $S_3=$180 nm, $S_4=$550 nm, $S_5=$180 nm, $S_6=$650 nm, and $S_7=$180 nm. (d) Emission efficiency of the antenna as a function of the width of its grating teeth. (e) Simulated far-field emission profile of the antenna.}
\label{fig:antenna}
\end{figure}

We employ a similar nanoantenna design used in~\cite{sun2013large,fatemi2018high} as the radiating elements of our PIC phased array. As illustrated in Fig.~\ref{fig:antenna}(a), the design features a 4~\textmu m taper connecting the PIC 470~nm wide silicon waveguide to the 2~\textmu m wide antenna. Fig.~\ref{fig:antenna}(b) illustrates a cross-section of the structure and highlights that it is buried between two thick layers of oxide. Fig.~\ref{fig:antenna}(c) displays some of the finer features of the antenna in the grating region of the device. Finite-difference time-domain (FDTD) simulations were used to find the dimensions of the nanoantenna to enhance its emission efficiency and field of view. Based on the simulated results shown in Fig.~\ref{fig:antenna}(d) of radiation efficiency versus the antenna grating tooth width, we used a 180~nm tooth width in our structure. The simulated far-field radiation pattern of the final nanoantenna design is displayed in Fig.~\ref{fig:antenna}(e) and can also be found in the main text.

\subsection{Near-field imaging resonator characterization}

\begin{figure}[t]
\centering\includegraphics[width=0.7\columnwidth]{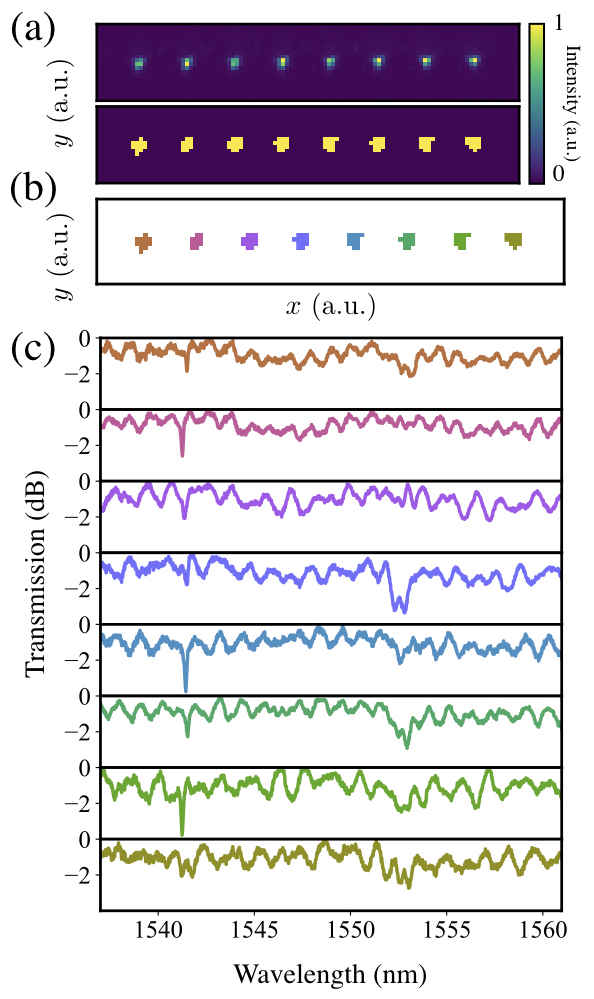}
\caption{\textbf{Extraction of the resonator transmission spectra from near-field images.} (a) Raw (top) and threshold-ed (bottom) versions of the near-field images of the light emitted from the phased-array. A threshold value of 0.1 was employed in the illustrated results. (b) Bins attributed to each nano-antenna obtained from the threshold-ed image in (a). (c) Resonator transmission spectra obtained based on the bin assignment shown in (b). The spectra are colored based on their bin assignment.}
\label{fig:imgProc}
\end{figure}

In order to accelerate the process of finding our phase shifter resonances, which is required for our device's calibration, we employ an image-based resonance characterization method. The latter consists of taking a sequence of images of the phased array near-field while sweeping over the optical wavelength used in our experiment. The collected images are then averaged and pixels at which an average intensity exceeds a given threshold are identified. These pixels are then clustered in bins based on their spatial proximity such that each bin corresponds to the near-field emitted by one of the phased array nanoantennas. These bins are then used to extract the total intensity of the field emitted by each antenna for all images collected during the wavelength sweep. The resulting data is then employed to reconstruct the emission spectrum of the device's resonator phase shifters. Sample results attributed to the steps taken in this approach can be found in Fig.~\ref{fig:imgProc}.

\subsection{Far-field imaging resonator characterization}

As mentioned in the main text, we can characterize the resonance of each array radiating element by performing a wavelength sweep of the array's far-field. The swept quantity consists of the mean square difference between images where a center, $V_0$, and a shifted voltage, $V_0+\delta V$, are applied onto the considered resonator, which can be mathematically expressed as
\begin{equation}
    \label{eq:ffSpectrum}
    f(\lambda) = \sum_{i,j} (I_{i,j}(V=V_0,\lambda)-I_{i,j}(V=V_0+\delta V,\lambda))^2,
\end{equation}
where $\lambda$ is the considered wavelength, $I$ denotes a collected image, the $i,j$ indices respectively indicate the image pixel coordinates, $V$ consists of the applied voltage, and $V_0$ and $V_0+\delta V$ are the center and shifted voltages applied to the device, respectively. Hereafter along with in the main text, we refer to $f(\lambda)$ as the squared differential far-field intensity (SDFI).

For a periodic phased array, one can model $f(\lambda)$ through the use of Fourier optics~\cite{born1999principles}. Namely, the far-field diffraction pattern of the phased array is expected to correspond to the result of the Fraunhoffer diffraction integral of the phased array's near field. For individual radiating elements defined by a near-field pattern $u_0(x,y,z)$, assuming that the elements are arranged along the $x$ direction and emitting along the $z$ direction, then the far-field diffraction pattern of the phased array is given by
\begin{equation}
    u(x,y,z) \propto \mathcal{U}_0 \left(\frac{kx}{z},\frac{ky}{z}\right)A(x,z),
\end{equation}
where $\mathcal{U}_0$ is the Fourier transform of $u_0$, $k$ is the radiation wavevector, and $A(x,z)$ is the array factor. When all elements are radiating in phase and are arranged periodically, this factor is given by:
\begin{equation}
    A(x,z) = \sum_{m=0}^{N-1} \exp{(j m k \Lambda x/z)},
\end{equation}
where $N$ is the number of emitting elements and $\Lambda$ is the periodicity of the array.

\begin{figure}[htbp]
\centering\includegraphics[width=\columnwidth]{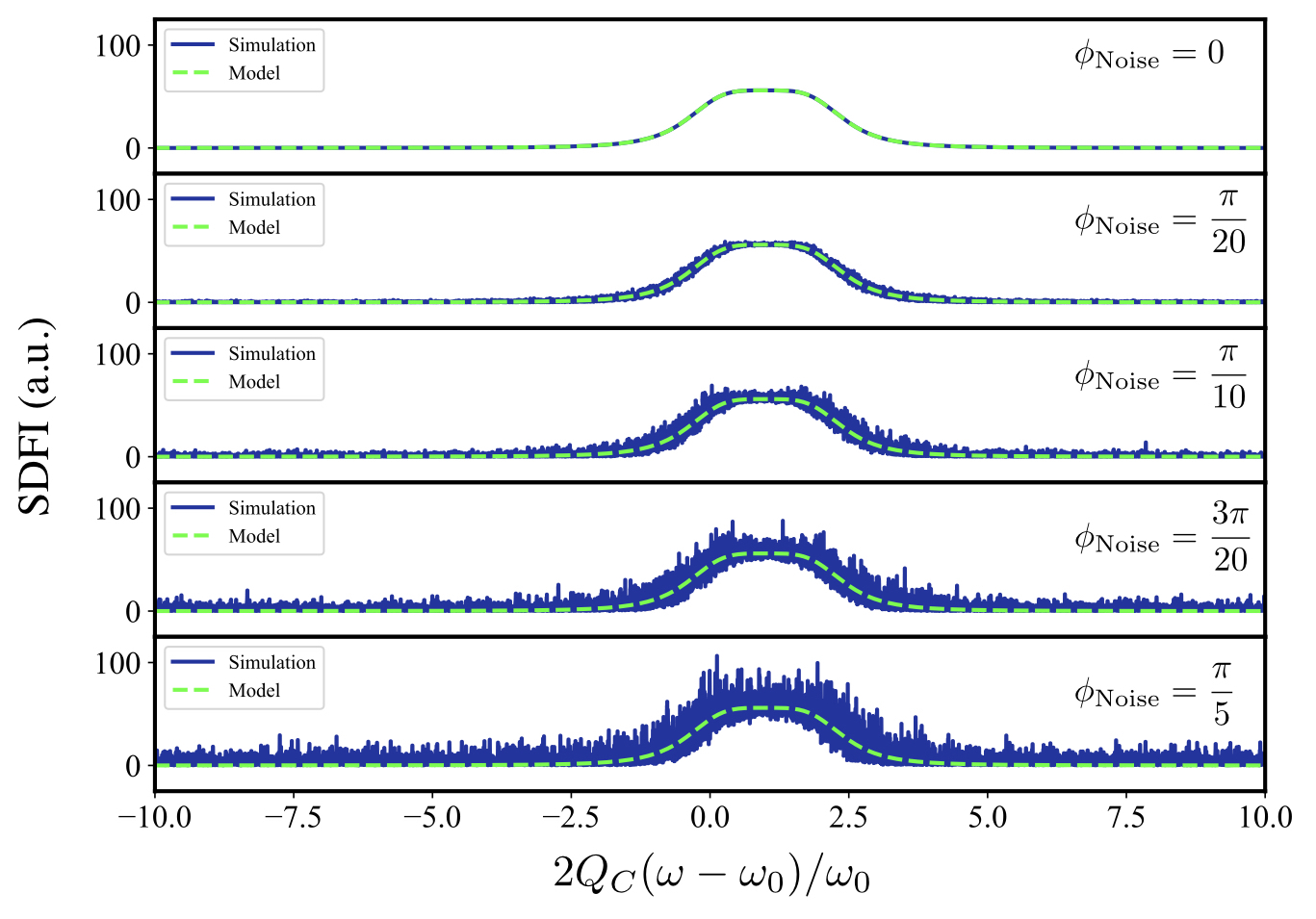}
\caption{\textbf{Simulated SDFI spectral response of a phased array under the presence of environmental perturbations.} The simulated far-field response accounts for variations in the transmissive properties of other resonators while shifting the resonance of the examined $1^\text{st}$ resonator in the phased array by $\delta\omega=\text{FWHM}$. The plotted curves are attributed to applied random phase noise of varying amplitude to the device phase shifters.}
\label{fig:simCurves}
\end{figure}

Consider the case where the $n^{\text{th}}$ element of the array has a phase offsetted by $\Delta \phi$ due to a voltage applied to the device's heater. The array factor then becomes
\begin{equation}
\begin{split}
    \label{eq:arrayPhaseRaw}
    A(\alpha &= k\Lambda x/ z, \Delta\phi) = \sum_{m=0}^{N-1} \exp{(j \, m \alpha + \delta_{mn}\Delta\phi)}\\
    & = \sum_{m=0}^{n-2} e^{j \, m \alpha} + e^{j \, (n-1) \alpha + \Delta \phi} + e^{j n \alpha} \sum_{m=0}^{N-n-1} e^{j \, m \alpha}\\
    & = \frac{1-e^{j(n-1)\alpha}}{1-e^{j\alpha}} + e^{j \, (n-1) \alpha + \Delta \phi} + \frac{e^{j n\alpha}-e^{j N\alpha}}{1-e^{j\alpha}},
\end{split}
\end{equation}
where $\delta_{mn}$ is the Kronecker delta function. We therefore expect the far-field intensity pattern formed by such a phased array to be given by
\begin{equation}
    I(x,y,z,\Delta\phi) \propto \left\lvert \mathcal{U}_0(\frac{\alpha}{\Lambda},\frac{ky}{z})\right\rvert^2  \left\lvert A(\alpha,\Delta\phi)\right\rvert^2,
\end{equation}
hence, following a few algebraic maneuvers, the difference between the two considered intensity patterns should be given by
\begin{equation}
\begin{split}
   I(x,y,z,\Delta\Phi) &- I(x,y,z,0) \propto \left\lvert  A(\alpha,\Delta\phi)\right\rvert^2 -\left\lvert A(\alpha,0)\right\rvert^2 \\
    =&\frac{4\sin{(\Delta\phi/2)}}{\sin{(\alpha/2)}}\left[\sin\left(\frac{\alpha}{2}\right)\sin\left(\frac{\Delta\phi}{2}\right) \right.\\ 
   &\left.- \sin\left(\frac{N \alpha}{2}\right)\sin\left(\frac{\Delta\phi-(N+1-2n)\alpha}{2}\right) \right]
\end{split}
\end{equation}
To obtain the SDFI spectrum given in Eq.~(\ref{eq:ffSpectrum}), the above expression must be squared and summed over all positions. In our case, the features of our far-field attributed to the array factor are much finer than those related to the far-field pattern of an individual antenna. For this reason, one may approximate the summation over all positions as an average over the $\alpha$ variable. The periodicity of the array factor ensures that this average can be taken from $-\pi$ to $+\pi$. Taking this average yields the expression provided below
\begin{equation}
\begin{split}
 f(\lambda) = 4 & \sin^2\left({\Delta\phi(\lambda)}/{2}\right) [2 \lvert N\rvert  +(2\lvert1+N-2n\rvert  \\
&- \lvert1+2N-2n\rvert-\lvert1-2n\rvert)\cos\left(\Delta\phi(\lambda)\right)\\
 &+4(1-\lvert1+N-n\rvert - \lvert n\rvert +  \lvert n-N \rvert \\
 &+ 2\lvert 1-n \rvert )\sin^2\left({\Delta\phi(\lambda)}/{2}\right)].
 \end{split}
\end{equation}
The above expression can then be simplified by recalling that $N>0$ and that $1\leq n \leq N$. Upon such consideration, $f(\lambda)$ can be more compactly written as
\begin{equation}
\label{eq:SDFI}
\begin{split}
 f(\lambda) = 16 & \sin^2\left(\frac{\Delta\phi(\lambda)}{2}\right) \left[(N-1)\sin^2\left(\frac{\Delta\phi(\lambda)}{2}\right)\right.\\
 &\left.+\left\lvert\frac{2n-N-1}{2}\right\rvert\cos{(\Delta\phi(\lambda))}\right].
 \end{split}
\end{equation}
Note that from symmetry, $f(\lambda)$ remains the same regardless of from which side of the array $n$ is labeled.
\begin{figure}[b]
\centering\includegraphics[width=\columnwidth]{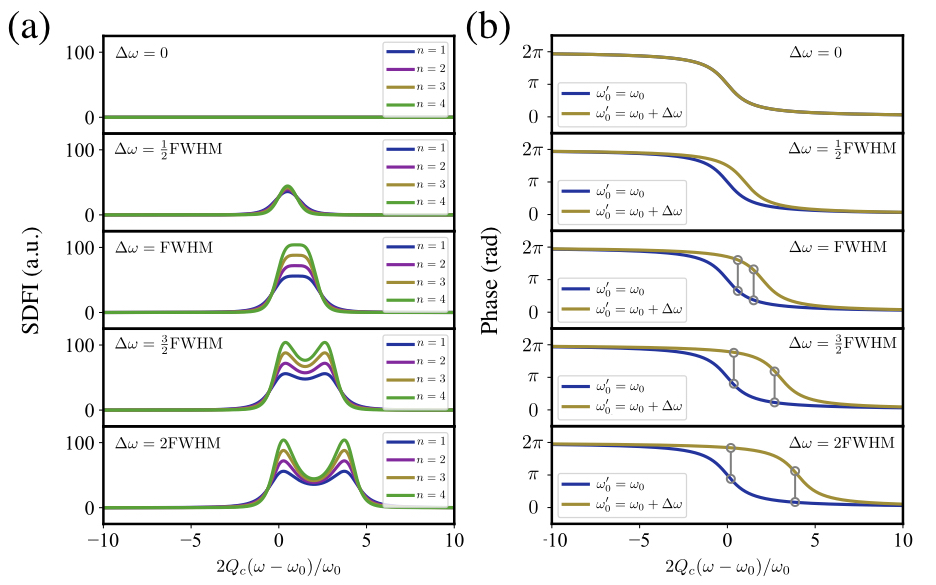}
\caption{\textbf{Modeled SDFI spectral response of a phased array to resonance shifts in a single resonator} (a) Modeled SDFI spectrum $f(\omega)$ attributed to various resonance shifts $\delta \omega$ applied to the first to the fourth emitting elements in the phased array. (b) Corresponding phase transmission spectra of the resonator for shifted ($\omega_0'=\omega_0$) and non-shifted ($\omega_0'=\omega_0 + \delta \omega$) resonant wavelengths, where wavelengths experiencing a $\pi$ phase shift upon the resonance shift are labeled. Resonant frequency shifts $\delta\omega$ are expressed in terms of the full-width half-maximum of the resonator's transmission spectrum, which, for over-coupled resonators, consists of $\text{FWHM} = \omega_0/Q_c$. Note that from both symmetry and Eq.~(\ref{eq:SDFI}), we expect curves associated with $n$ and $N-1-n$ to be identical.}
\label{fig:theoCurves}
\end{figure}
Plots of the SDFI in frequency space, $f(\omega)$, attributed to shifting the resonances of various heaters in a 1x8 phased array can be found in Fig.~\ref{fig:theoCurves}. As expected, the largest changes in the far-field spectra occur for wavelengths that experience a $\pi$ phase shift upon shifting the resonance of the device phase shifters.
Our modeled $f(\omega)$ accurately reproduces a simulated version of it taking into account the conditions of our experiment provided that the phase shifters whose phases are not varied while the examined phase shifter shifts its resonance, regardless of whether all phase shifters emit in phase. However, as illustrated in Fig.~\ref{fig:simCurves}, noise begins to affect $f(\omega)$ when fluctuations in the transmission of the other phase shifters occur while the resonance of the examined radiating element is shifted. Such fluctuations can be attributed
to environmental perturbations in our apparatus and are also observed in our experimental curves as shown in the main text.

\end{document}